\documentclass[%
reprint,
 aps,
]{revtex4-1}

\usepackage{graphicx}
\usepackage{dcolumn}
\usepackage{bm}
\usepackage{amsmath}
\usepackage[utf8]{inputenc}
\usepackage[T1]{fontenc}
\usepackage{mathptmx}
\usepackage{textcomp}
\usepackage{gensymb}
\usepackage{tensor}
\usepackage{xcolor}
\usepackage{braket}

\usepackage[colorlinks=true,linkcolor=red,citecolor=blue,urlcolor=blue]{hyperref}

\newcommand{\Temple}{Department of Physics, Temple University, Philadelphia, Pennsylvania 19122, USA}
\newcommand{\UW}{Department of Physics,  University of Washington, Seattle, Washington 98195, USA}

\begin{document}

\title{Neutral Atoms in Optical Tweezers as Messenger Qubits for Scaling up a Trapped Ion Quantum Computer}

\author{Svetlana Kotochigova}
\affiliation{\Temple}
\author{Subhadeep Gupta}
\affiliation{\UW}
\author{Boris B. Blinov}
\affiliation{\UW}

\date{\today}

\begin{abstract}
We propose to combine neutral atom and trapped ion qubits in one scalable modular architecture that uses shuttling of individual neutral atoms in optical tweezers to realize atomic interconnects between trapped ion quantum registers. These interconnects are \textit{deterministic}, and thus may be performed \textit{on-demand}. The proposed protocol is as follows: a tweezer-trapped neutral atom qubit is brought close to a
trapped ion in an ion chain serving as a module of a larger quantum computer, and an entangling gate is performed between the two qubits. Then the neutral atom is quickly moved to another, nearby trapped ion chain in the same modular ion trap and entangled with an ion in that chain, thus entangling the two separate ion chains. The optical dipole potential of the tweezer beam for the neutral atom does not measurably affect the trapped ions, while the RF ion trap does not affect the neutral atom. With realistic tweezer trap parameters, the neutral atom can be moved over millimeter scale distance in a few tens of microseconds, thus enabling a remote entanglement generation rate of over $10^3$/s
even with very modest assumptions for the atom-ion quantum gate speed, and possibly up to $10^4$/s, which is two orders of magnitude higher than the current state-of-the-art with photonic interconnects. 

\end{abstract}

\maketitle

Scaling up a quantum computer is one of the major challenges for the trapped ion qubit technology. Two major contenders are the modular architecture featuring small trapped ion qubit registers connected via photonic links, known as MUSIQC (Modular Universal trapped Ion Quantum Computer)~\cite{MUSIQC} and the quantum charged-coupled device (QCCD) architecture with ions shuttled between different trapping zones in a complex interconnected trap structure~\cite{Kielpinski2002}. Other proposals for scaling up include the large 2-dimensional trapped ion crystals~\cite{Guo2024} and the laser-free architecture with microwave qubit gates and magnetic field gradients~\cite{Mintert2001}.
However, scaling these systems beyond a few hundred qubits while maintaining high quantum gate fidelity is challenging. In the MUSIQC architecture, the photonic interconnects are created probabilistically, and the rate at which these interconnects are created is limited by the ability to collect and detect single photons from the ions. This necessitates the generation of these interconnects off-line for later use. In the QCCD architecture, shuttling of ions creates time and resources overheads, and the ion chain splitting, ion shuttling and chain merging create heating~\cite{Pino2021}. On the other hand, arrays of {\it individually movable} single neutral atoms in optical tweezer traps have recently enjoyed a very rapid and successful development as a tool for building and scaling up quantum information processing systems \cite{Bluvstein2022, Madjarov2020, Graham2023, mane24}. 

\begin{figure}
\includegraphics [width=0.49\textwidth]{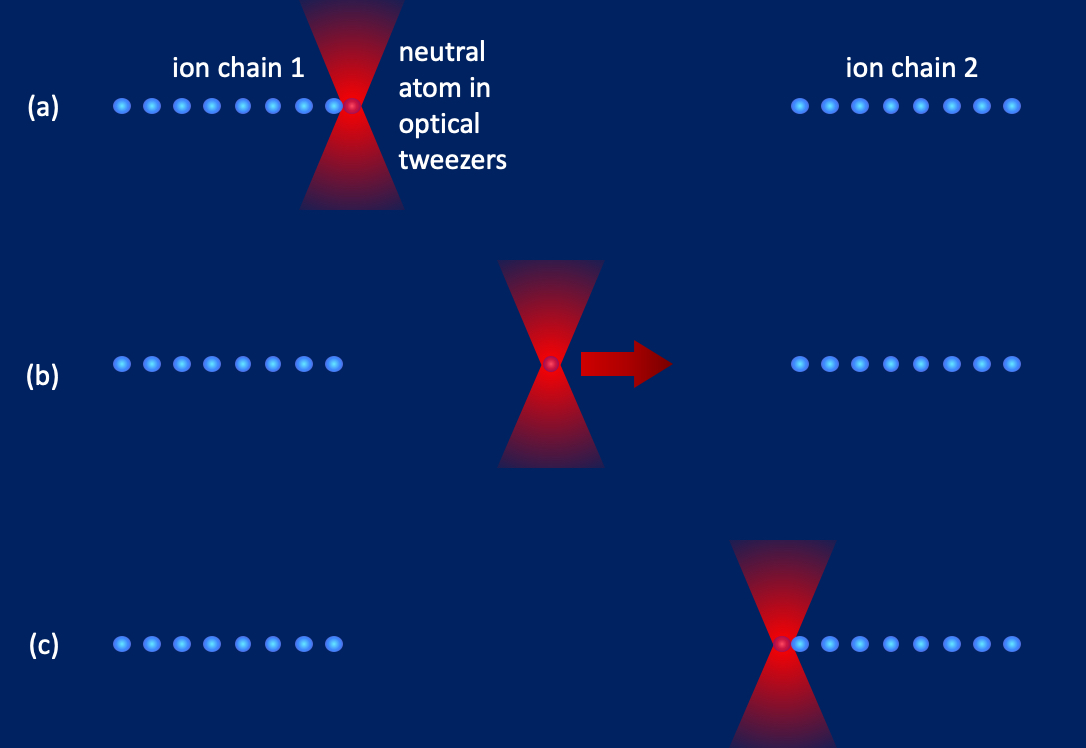}
  \caption{The conceptual framework of the remote ion entanglement using trapped neutral atoms as ``flying" qubits. (a) A neutral atom in an optical tweezer trap is entangled with the right-most ion in ion chain 1. (b) The neutral atom is then rapidly transported towards the second ion string in a separate potential well. (c) The tweezer-trapped neutral atom is brought near the left-most ion in ion string 2 in order to entangle them, thus entangling ions in the two separate strings.}
  \label{Fig:concept}
\end{figure}

Here we propose a novel architecture for scaling up quantum information processors based on trapped atomic ions. This modular design uses individual neutral atoms trapped in optical tweezer traps as messenger qubits to {\it deterministically} distribute entanglement between trapped-ion quantum computing modules. Our proposed protocol is is depicted in Fig.~\ref{Fig:concept}. A tweezer-trapped atom is brought close to a trapped ion in a relatively small ion chain serving as a module of a larger quantum computer. An entangled state is then generated between the two qubits via, for example, a controlled spin-state-dependent collision, or a Rydberg-type laser-assisted gate. The neutral atom is then rapidly moved by translating the optical tweezer to another trapped ion chain in the same vacuum system. A second entanglement operation is then performed between the atom and an ion in this second chain, thus connecting the two separate chains. The optical dipole potential for the tweezer trap can be designed to not measurably affect the trapped ions, while the RF ion trapping potential will not affect the neutral atom. Thus, the neutral atom can be a versatile tool for entangling trapped ions both between distant ion chains within the same vacuum system, and within an individual ion chain. This versatility substantially reduces time and heating overheads in our proposed protocol compared to that encountered in the QCCD architecture while the deterministic entanglement generation addresses the drawback of the MUSIQC architecture.

\begin{figure}
\includegraphics[width=0.49\textwidth]{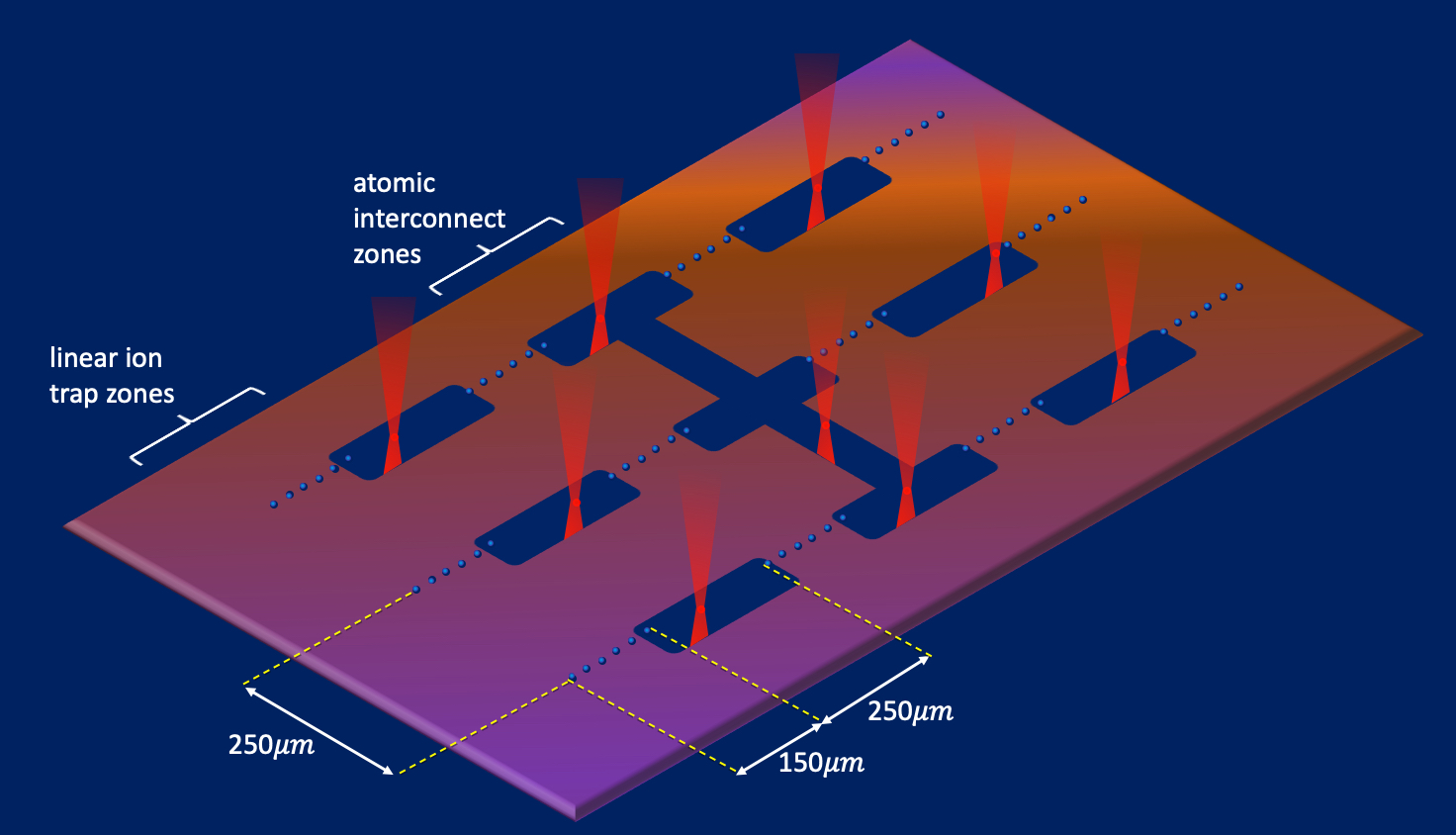}
\caption{A concept for scaling up the system. A chip ion trap with multiple trapping zones defined by RF and DC electrodes (not shown for clarity) contains several linear ion chains with several tens of ion qubits each.  Multiple through holes in the chip allow for optical tweezer access, thus allowing atomic interconnects. Assuming 30-ion chains with 5 $\mu$m spacing between ions and 250 $\mu$m separation between the trapping zones, up to 20 ion trap zones can fit within the 1.2~mm diameter field of view of the high-NA lens used for the optical tweezers. Atoms may be loaded into individual tweezers from a larger optical dipole trap formed by a laser beam parallel to the ion trap chip surface (not shown).}
\label{fig:scaling}
\end{figure}

We now turn to the scalability of the proposed architecture. Scaling up is possible in a chip trap setup as depicted in Fig.~\ref{fig:scaling}. The multiple linear chains of ions trapped above the chip are located in the focal plane of the high-NA lens used for the optical tweezers and represent the individual quantum computer modules of this architecture. Multiple tweezers laser beams are used to move the trapped atom messenger qubits between these ion trap modules to achieve a fully interconnected quantum processor. Assuming a typical 1.2 mm diameter field-of-view (FOV) high-NA lens, up to 20 such ion chains with 30 ion qubits per chain can fit within the FOV, resulting in a 600-qubit device. Further scaling up may be possible through the engineering of larger FOV lenses, or by integrating the tweezers beam optics into the ion trap chip.

As perhaps with any existing or proposed trapped-ion quantum computer architecture, scaling beyond a single chip with order 1000 qubits may have to be done via photonic interconnects~\cite{MUSIQC}. In the presented system, we can envision series of ions in separate trapping zones used specifically for the photonic interlink generation. These photonic entanglement zones are separated by hundreds of microns from the local quantum logic zones, and thus, single trapped ion qubit species can be used for both tasks, avoiding the additional complication of multiple ion qubit species. Alternatively, with optical and mechanical design, the messenger neutral atom qubits may be moved between distant ion trap chips within the same vacuum system, thus avoiding the use of probabilistic ion-photon entanglement altogether.

We now consider the timescale for logic operations in this protocol, which can be divided into the timescale for the entangling atom-ion gate and the timescale for coherent transport of the messenger qubit. For the protocol shown in Fig 1, an important limitation to the generation of entanglement between separate ion chains in the proposed architecture is the timescale over which a tweezer-trapped atom can be coherently transported between the ions located in two ion trapping zones. The optical tweezer beam is transported by rapidly sweeping the drive frequency of an acousto-optic deflector (AOD). The resulting deflection angle change is transferred into translation of the laser beam waist with the focusing objective lens. Using controlled radiofrequency drive waveforms programmed with an Arbitrary Waveform Generator, the desired velocity and acceleration profiles can be achieved. Recently, fast coherent transport of a tweezer-trapped $^{87}$Rb atom was demonstrated~\cite{Bluvstein2022}, with an average speed of 0.55$\,\mu$m$/\mu$s. The main limitation of the speed of the transfer was the mass of the atom, which in turn limited the trap frequency. Faster implementation will be possible with a lighter atom, and for the rest of this paper we will consider the lightest alkali atom, lithium. 

Tweezer trapping of individual ${^6}$Li atoms has recently been demonstrated \cite{blodgett2023}. Using the similar beam waist of $1\,\mu$m and tweezer laser wavelength of 1064~nm as used in that work, we arrive at a 10~mK trap depth for 250~mW laser power, implying radial (axial) trap frequency of $\omega_{{\rm Li},r(z)}=2\pi \times 1.2 (0.28)\,$MHz. These high trap frequencies would result in an order of magnitude reduction in the timescale of tweezer beam translation as compared to~\cite{Bluvstein2022}. One can thus expect to coherently transport a Li atom between individual ion trapping zones in a segmented linear ion trap, about 250$\mu$m, in under 50$\mu$s. With optimized optical geometries, as proposed in \cite{premawardhana2024}, acceleration of up to even $10^5$m/s$^2$ may be possible, corresponding to translation of 250$\mu$m in less than a microsecond. 

While the discussion so far has been limited to tweezer beam translation in the transverse direction (as depicted in Fig.\ref{Fig:concept}) using an AOD, axial translation of the trap (perpendicular to the focal plane of the tweezer lens) is also possible using a varifocal (focus tunable) lens \cite{kang2020}, which would liberate the architecture from the constraint of the tweezer lens FOV. Focus changes in few tens of microseconds may be possible with such devices \cite{mermillod08}, allowing fast transport and providing another avenue for up-scaling of the proposed architecture. 

We now turn to a central feature of the proposed architecture: the entangling two-qubit gate between a trapped neutral atom and a trapped ion. We consider two options: a collisional gate and a Rydberg gate. The collisional gate is based on spin-dependent phase accumulated during the controlled ultracold collision between the atom and the ion~\cite{Doerk2010}, while the Rydberg gate is enabled by the long-range dipole-dipole interaction or the van-der Waals interaction between the atom and the ion in highly excited Rydberg states~\cite{Saffman2010,Levine2019,Zhang2020}. 

The collisional gate is attractive due to its simplicity: no additional lasers are needed, and the gate itself is performed by bringing the atom and the ion close together for a specific amount of time and then separating them. Such a gate has been proposed in~\cite{Doerk2010}, where a $^{87}$Rb - $^{135}$Ba$^+$ qubit pair was theoretically investigated. Gate fidelity of up to 99.99\% was shown to be achievable within a $t_{\rm gate}=300~\mu$s gate time. Similar gate times should be possible with other atom-ion pairs. Furthermore, the use of an atom-ion magnetic Feshbach resonance \cite{weckesser2021} should allow for even faster collisional gates.

\begin{figure}
  \includegraphics[width=0.49\textwidth]{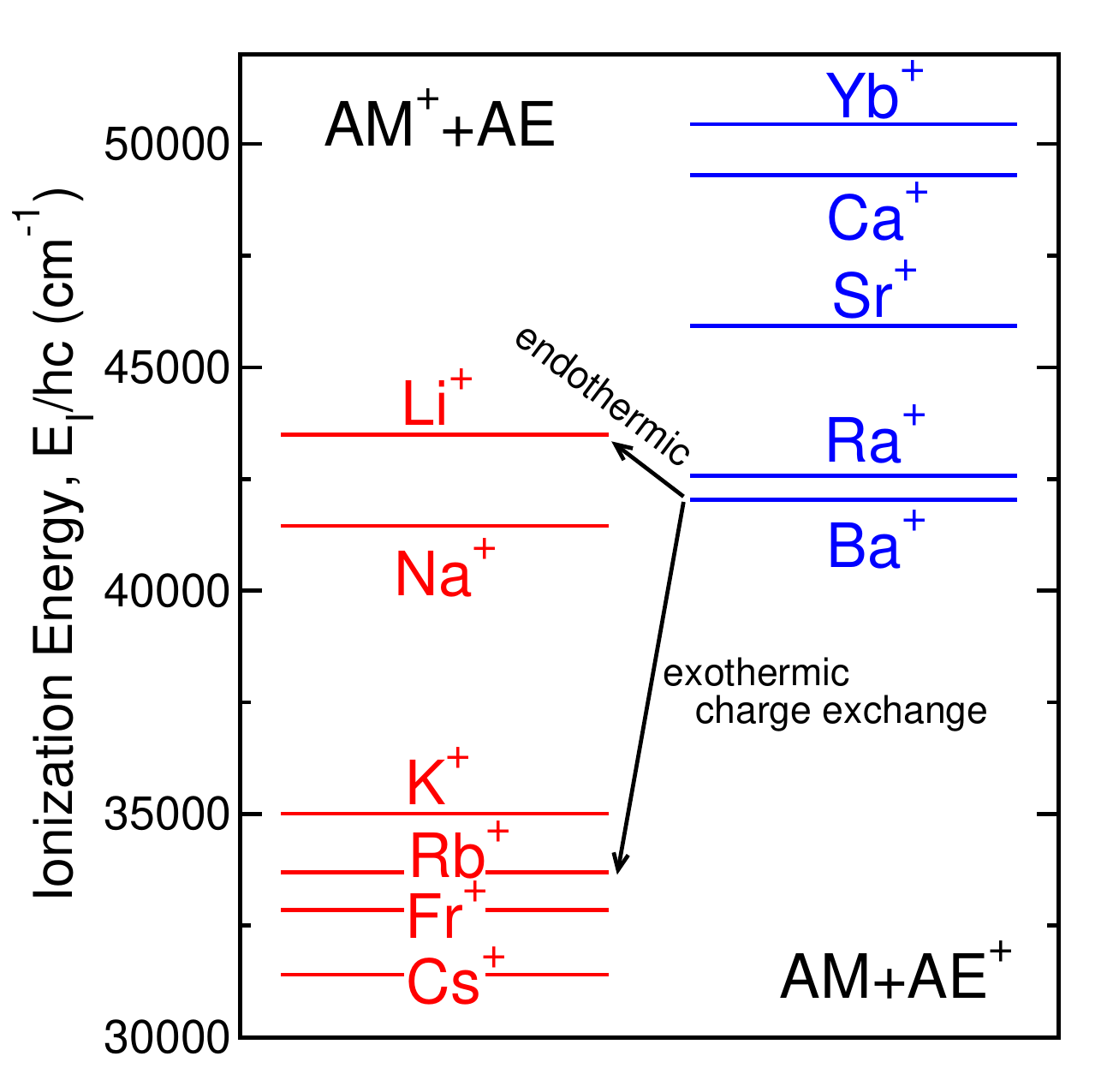}
  \caption{Ionization energies of alkali-metal (AM, red lines) and alkaline-earth and ytterbium (AE, blue lines) atoms relative to the corresponding neutral ground-state atom.
The ionization energies of Be and Mg are greater than $hc\times 55\,000$ cm$^{-1}$, and are not shown. The graph can be used to infer whether the charge exchange reaction between an ultracold neutral alkali-metal atom and a singly-charged alkaline-earth ion is endothermic or exothermic.  For example, for ${\rm Li}+{\rm Ba}^{+}$ and ${\rm Rb+Ba}^{+}$ the reactions are endothermic and exothermic, respectively.}
  \label{fig:ionization}
\end{figure}

An important consideration for the collisional gate is the stability of the system against charge-exchange collisions. In Fig.~\ref{fig:ionization}, we show the ionization energies of the alkali atoms that are good neutral atom qubit candidates as well as those of the alkali earth atoms (plus ytterbium), whose singly-charged states are good trapped ion qubit candidates. We find that in a charge-exchange collision, the only endothermic, and therefore energetically unfavorable ion-atom pairs are Ra$^+$-Li and Ba$^+$-Li, with the latter having the larger energy barrier. Additionally, lithium is the lightest of the alkalis, and it is therefore the best candidate for a high-rate transport of the optical-tweezers-trapped atom, as discussed above. For all these reasons, we believe that the Ba$^+$-Li system may be the perfect candidate for an atom-ion quantum computing architecture with collisional gates.

\begin{figure}
\includegraphics[width=0.47\textwidth]{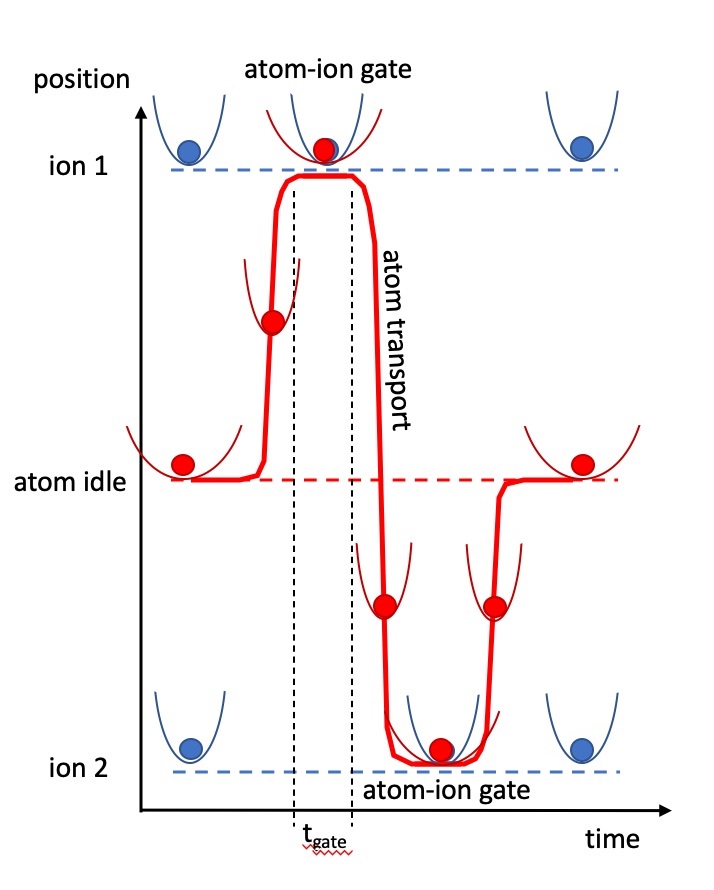}
\caption{Position vs. time graph for the two communication ions and the messenger neutral atom. The blue dashed lines indicate the positions of the ions, which do not change during the execution of the protocol. The dashed red line is the "idle" position of the neutral atom, while the solid red line is the neutral atom position during the protocol execution, including the periods of rapid transfer (high slope segments) and the atom-ion collisional gate periods (near-zero slope segments). The relative durations of these periods are not to scale as the individual atom-ion gate duration may be large compared to the atom transport times. Additionally, the optical tweezers trap strength may have to be adjusted between the rapid movement (tighter trap) and the gate execution (weaker trap).}
\label{fig:gate}
\end{figure}

The envisioned collisional gate procedure is illustrated in Fig.~\ref{fig:gate}. A neutral atom messenger qubit is held initially at an "idle" position away from the trapped ions. Once an interconnect needs to be established between the ionic qubits, the atom is quickly brought close to one of the ions. When the atom has approached the vicinity of the ion, the spatial overlap between the two qubits is optimized and the effect of the tweezer beam on the ion is minimized by dynamically reducing the tweezer beam power. We note that in practice, the gate procedure will be optimized by theoretical and experimental investigation of the effect of the tweezer potential on the ion-atom collision process. The total time to establish entanglement between separate trapped ion chains can be estimated as twice the gate time plus the transport time, or $2\times300~\mu$s~+~$50~\mu$s$=650~\mu$s, using the most conservative values discussed earlier. The resulting entanglement generation rate is about $1.5\cdot10^3$/s which is nearly an order of magnitude higher than the state-of-the-art rate using photonic interconnects~\cite{Jameson2024}.

\begin{figure}
\includegraphics[width=0.47\textwidth]{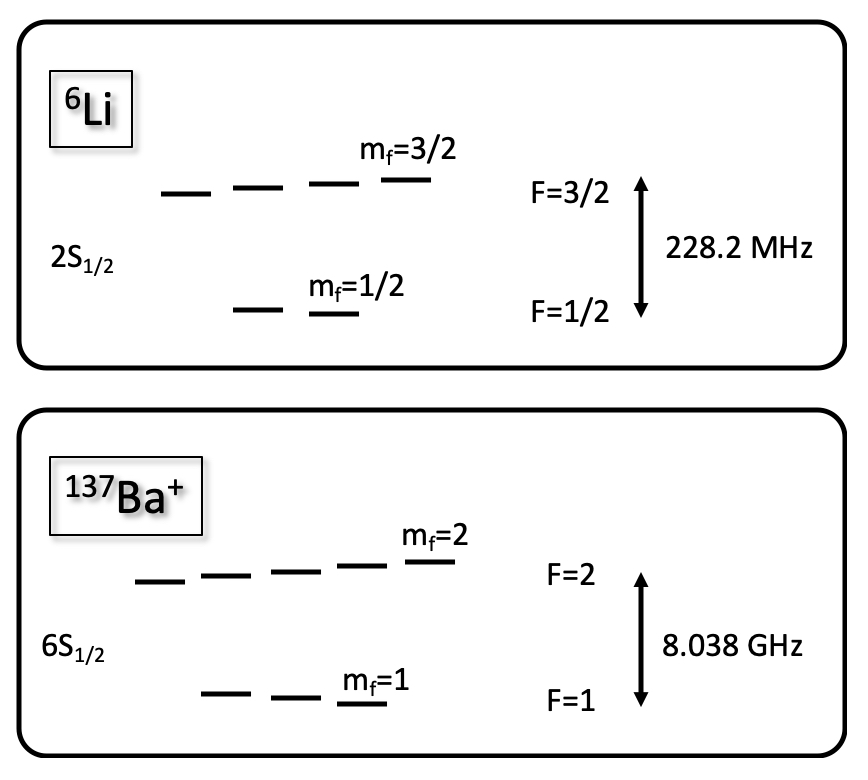}
\caption{Ground state hyperfine manifolds (not to scale) of $^{6}$Li (top) and $^{137}$Ba$^+$ (bottom) showing the relevant qubit states: $|F=1/2,m_F=1/2\rangle \equiv |0_a\rangle$ and $|F=3/2,m_F=3/2\rangle \equiv |1_a\rangle$ for lithium atom and $|F=1,m_F=1\rangle \equiv |0_i\rangle$ and $|F=2,m_F=2\rangle \equiv |1_i\rangle$ for barium ion.}
\label{fig:qubits}
\end{figure}

The qubit possibilities in the $^{137}$Ba$^{+}-^{6}$Li system involved in an individual ion-atom gate can be illustrated through the ground-state hyperfine structures of the two atoms (see Fig.~\ref{fig:qubits}). The ion-atom collisional gate~\cite{Doerk2010} can be represented as:
\begin{align*}
|0_i0_a\rangle &\rightarrow |0_i0_a\rangle \\
|0_i1_a\rangle &\rightarrow |0_i1_a\rangle \\
|1_i0_a\rangle &\rightarrow |1_i0_a\rangle \\
|1_i1_a\rangle &\rightarrow e^{-i\phi}|1_i1_a\rangle
\end{align*}
where the resultant phase $\phi$ from the collisional interaction needs to be equal to $\pi$ for a two-qubit controlled-phase (Cz) gate. Adapting the qubit states discussed in \cite{Doerk2010}, the choice of $^{137}$Ba$^+$ 6S$_{1/2}$ states $|F=1,m_F=1\rangle \equiv |0_i\rangle$ and $|F=2,m_F=2\rangle \equiv |1_i\rangle$ for the ionic qubits and $^{6}$Li 2S$_{1/2}$ states $|F=1/2,m_F=1/2\rangle \equiv |0_a\rangle$ and $|F=3/2,m_F=3/2\rangle \equiv |1_a\rangle$ for the atomic qubits will provide some protection from spin-changing collisions. Ba$^{+}$ also offers the possibility of an optical qubit via the 6S$_{1/2}$ and 5D$_{5/2}$ states. The availability of additional stable isotopes ($^{138}$Ba$^{+}$ and $^{7}$Li) provide further opportunities for controlling the collisional gate properties, including through ion-atom magnetic Feshbach resonances \cite{weckesser2021}.

The fidelity of the collisional gate will depend on additional factors, including inelastic losses due to spin-flips mentioned above. Ideally, the loss rates $K_{\rm loss}n$ should be orders of magnitude smaller than the rate $1/t_{\rm gate}$ at which the quantum phase gate is generated, where $K_{\rm loss}$ is the microscopic loss rate coefficient that depends on the atom-ion interaction potentials and $n$ is the relative number density of the Ba$^+$ and Li pair at their closest approach. Detailed calculations of the spin-flip loss rate coefficients for the Ba$^+$-Li collision show that this condition is not always met and requires a careful choice of pairs of hyperfine states of the $^6$Li atom \cite{Tiesinga2025}. Additionally, the cooling requirements for the collisional gate, such as cooling the trapped ion to near the ground state of the harmonic confining potential, as well as the exact atom-ion positioning requirements during the collisional gate execution are rather stringent. 

An alternative is to use the Rydberg gate of the type demonstrated both in the atom-atom~\cite{Levine2019} and the ion-ion~\cite{Zhang2020} systems. We note that there have been no demonstrations of a neutral atom-trapped ion Rydberg gate to date. In the neutral atom system, the van der Waals interaction is sufficiently strong to induce the Rydberg blockade when the atoms are excited to states with principal quantum number of order 50 within several microns~\cite{Saffman2010}. In the ion-ion system, the same interaction is much weaker, since its strength scales as $Z^{-6}$ with the net core charge $Z$, which is +2 for singly charged ions~\cite{Lebedev1998}. However, in the neutral atom-ion system the scaling is effectively $Z^{-3}$, so the van der Waals interaction should be sufficiently strong to enable a blockade-type gate at a separation of order 1~$\mu$m, which would be readily achievable in the proposed architecture. Thus, the tweezer-trapped atom would only need to be brought sufficiently close to the ion, but the optical dipole potential can stay high, and the tweezer beam focus does not have to overlap the trapped ion, significantly reducing both the spontaneous emission probability and the motional heating of the ion. Rydberg interactions can thus enable fast, \textit{motion-state-independent} gates between trapped ions and neutral atoms, considerably relaxing the requirements on trapped ion cooling and eliminating the need to cool to the motional ground state. The entangling operation procedure would be very similar to the one shown in Fig.~\ref{fig:gate}, but without the adjustment of the optical dipole trap strength during the gate, and without the requirement to spatially overlap the atom and the ion. The gate time may also be much shorter, in the sub-microsecond regime~\cite{Zhang2020}, thus enabling even higher remote entanglement generation rate of order $2\cdot10^4$/s. 

A notable drawback of the Rydberg gate is the necessity to introduce additional lasers to excite the Rydberg states. Rydberg excitation of Li atoms \cite{Hulet1983} can be performed using a two-photon process from 2S$_{1/2} \rightarrow$ 2P$_{3/2} \rightarrow$ nS with 671~nm and 350~nm laser light. For Ba$^+$, Rydberg excitation may be achieved by first shelving one of the qubit states to the metastable 5D$_{5/2}$ state with a 1762~nm laser~\cite{Dietrich2010}, followed by a two-photon process 5D$_{5/2} \rightarrow$ 7P$_{3/2} \rightarrow$ nS using 225~nm and 329~nm lasers.

In conclusion, we have proposed a novel architecture for scaling up a trapped-ion based quantum computer by using optical tweezer-trapped neutral atoms as messenger qubits to connect separate ion trapping zones within the same vacuum chamber. We discussed details of the atom transport and two potential candidate entangling gates. We have shown that the particular combination of trapped ion and neutral atom species, Ba$^+$Li, offers fast neutral atom transport and short entangling gate times, but other neutral atom-trapped ion combinations may also be considered. Scaling up to hundreds and thousands of qubits is within reach of existing technology, and further up-scaling should be possible. By combining the best of both atomic worlds we can pave the way to scaling the modular quantum processor arrays with this deterministic atomic qubit remote entanglement. The neutral atom qubit may also be used to entangle the trapped ion qubits \textit{within} the individual ionic modules, offering a new paradigm for trapped ion/trapped atom quantum computing architectures. 

\section*{Acknowledgments}
We would like to thank Eite Tiesinga
and Jacek K\l{}os for useful discussions. The work of S.K. was supported by the US Air Force Office of Scientific Research Grant No. FA9550-21-1-0153 and the National Science Foundation Grant No. PHY-2409425. The work of S.G. was supported by the US Air Force Office of Scientific Research Grant No. FA9550-22-1-0240. The work of B.B. was supported by the National Science Foundation Grant No. PHY-2308999 and the U.S. Department of Energy, Office of Science, Office of Basic Energy Sciences, Award No. DESC0020378.
\bibliography{KGB}

\end{document}